\documentclass[conference]{IEEEtran}
\IEEEoverridecommandlockouts
\usepackage{cite}
\usepackage{amsmath,amssymb,amsfonts}
\usepackage{algorithmic}
\usepackage{algorithm}
\usepackage{graphicx}
\usepackage{multirow}
\usepackage{textcomp}
\usepackage{xcolor}
\def\BibTeX{{\rm B\kern-.05em{\sc i\kern-.025em b}\kern-.08em
    T\kern-.1667em\lower.7ex\hbox{E}\kern-.125emX}}

\begin{document}

\title{Enhance the performance of navigation:\\ A two-stage machine learning approach} 

\author{
Yimin Fan, Zhiyuan Wang, Yuanpeng Lin, Haisheng Tan\\
University of Science and Technology of China, Hefei, Anhui, P.R. China

}

\maketitle

\begin{abstract}
Real time traffic navigation is an  important capability in smart transportation technologies, which has been extensively studied these years. Due to the vast development of edge devices, collecting real time traffic data is no longer a problem. However, real traffic navigation is still considered to be a particularly challenging problem because of the time-varying patterns of the traffic flow and unpredictable accidents/congestion. To give accurate and reliable navigation results, predicting the future traffic flow(speed,congestion,volume,etc) in a fast and accurate way is of great importance. In this paper, we adopt the ideas of ensemble learning and develop a two-stage machine learning model to give accurate navigation results. We model the traffic flow as a time series and apply XGBoost algorithm to get accurate predictions on future traffic conditions(1st stage). We then apply the Top K Dijkstra algorithm to find a set of shortest paths from the give start point to the destination as the candidates of the output optimal path. With the prediction results in the 1st stage, we find one optimal path from the candidates as the output of the navigation algorithm. We show that our navigation algorithm can be greatly improved via EOPF(Enhanced Optimal Path Finding), which is based on neural network(2nd stage). We show that our method can be over 7\% better than the method without EOPF in many situations, which indicates the effectiveness of our model.
\end{abstract}


\section{Introduction}
\subsection{Background}
Nowadays, with the increasing population in big cities, the traffic systems are under serious pressure. People are hoping for better driving experience in cities as congestion problems are becoming more and more serious. Vast development of smart traffic technologies have offered several solutions to traffic problems. Our work focuses on the modern traffic navigation system. Traditional navigation algorithms merely focus on the current traffic topology and traffic data, which does not work well when the traffic condition is highly non-stationary and time varying. Though there are many works focusing on the traffic prediction problems, they do not show the performance of their algorithms when used in navigation systems. In this paper, we use the Top K Dijkstra algorithm as the basis of our navigation system. The system will find K shortest paths from the start point to the destination effectively. We develop a two-stage machine learning method to enhance the performance of our navigation system. In the first stage, we use the XGBoost algorithm, which is very effective in many machine learning tasks, to predict the future traffic conditions, including the future traffic speed, the probability of congestion, the stability and vulnerability of traffic conditions. This approach can provide extra information to our navigation system. Additionally, we propose two factors to quantify the usability of our algorithm and show when it works or doesn't work. In the second stage, we develop the EOPF(enhanced optimal path finding) algorithm, which is based on neural network. The algorithm can learn from our history predictions and navigation results to provide a better navigation result for the users.
 \subsection{Contributions}
Our paper makes contributions mainly in the following aspects:
\begin{itemize}
    \item apply the XGBoost algorithm in traffic prediction problems and achieve great performance.
    \item propose two factors to quantify the usability of traffic prediction model.
    \item quantify the congestion probability and stability of traffic conditions
    \item propose a two-stage machine learning approach to enhance the performance of navigation systems.
\end{itemize}
\section{Related Work}
Many existing literature focuses on traffic prediction and navigation problems, but few works consider the two problems as a whole and evaluate their relationship. 

Works focusing on traffic prediction mainly have two streams: the model-based approaches and data-driven approaches. The model-based approaches mostly focus on using comprehensive system modeling for the traffic topology, establishing differential equations or finding quantitative relationship between speed,volume and topology. Typical examples include \cite{b2},\cite{b3},\cite{b4}. The model-based approaches are often used for analysis rather than prediction, because their performances are relatively poor in comparison with data-driven approaches. One reason for that is the traffic condition is affected by many factors and it is impossible to establish a model for the whole system. One advantage of model-based approaches is that they are highly explainable and many useful conclusions(usually with domain knowledge) can be drawn from the results of these methods. The data-driven approaches are very popular these years. They do not try to quantify or give "closed-form" solution to the physical properties or dynamic behavior of traffic conditions. Instead, they use statistics or machine learning methods and try to capture the hidden patterns in history traffic data. They use what they capture from history data to predict the future traffic conditions. Typical examples include the use of multivariate linear regression \cite{b5}, autoregressive integrated moving average model\cite{b6}, support vector regression\cite{b7}, deep neural network\cite{b8}, long short term memory network\cite{b9},\cite{b10} and graph convolution network\cite{b11}.

Works focusing on traffic navigation mainly deal with the following problems. Fast and efficient algorithm for shortest path finding in graph is the basis of traffic navigation. Dijkstra algorithm is the most famous algorithm for that problem. Contraction Hierarchies\cite{b12} is another powerful algorithm for shortest path finding and it is extremely fast on large scale traffic network. Customizable route planning\cite{b13} is widely used because it computes shortest paths on continental road networks with arbitrary metrics. A* algorithm and its derivatives\cite{b14} are also useful in solving that problem. These years, traffic navigation problems in autonomous driving have been extensively studied\cite{b15},\cite{b16}. They use deep reinforcement learning to enhance the performance of path planners. These attempts may make traditional navigation algorithms more powerful in the future.
\section{Model Definition}
Our model basically contains two parts: the predict model and the navigation model. We will define and explain these two models in the following two subsections respectively. \subsection{Predict Model}
 Let $x^{i}_{t}$ be the traffic flow speed of road $i$ at time $t$. Let $f$ denote our model which maps input traffic condition to predictions on the future. Given the traffic flow speed at time $t,t+1,...,t+h$ on road $i$, where $h$ is called  \textit{window size}, we construct a state vector $V_{t:t+h}^{i}=(x^{i}_{t},x^{i}_{t+1},...,x^{i}_{t+h})$.
 
In our predict model, we mainly focus on the following problems. Given the speed of road $i$ in the past few hours/minutes(e.g.5 hours):
\begin{itemize}
    \item[a.] Predict the speed of road $i$ in the upcoming few hours/minutes(e.g.30 minutes).
    \item[b.] Predict the probability of congestion on road $i$ in the upcoming few hours/minutes.
    \item[c.] Predict the probability distribution of speed on road $i$ in the upcoming few hours/minutes.
\end{itemize}

In Problem a, We define the predict traffic flow speed of road $i$ at time $t$ as $\hat{X}^{i}_{t}$, and our problem can be written as:
\begin{equation}
\begin{split}
Min_{f}:Criterion(x^{i}_{t+d},\hat{x}^{i}_{t+d})\\
\textit{s.t.}\quad \hat{x}^{i}_{t+d}=f(V_{t:t+h}^{i})
\end{split}
\end{equation}

The $Criterion(\quad,\quad)$ in this problem can be defined as the squared error or absolute error, which gives measurements of the distance between our prediction and the real speed(which is unknown when we make predictions).

In Problem b, we say that the road is in congestion condition when its speed is below $p$ quantile of history speed. $p$ can be set differently in different cities and different traffic topology. In a city where congestion is very likely to happen, we set $p$=20 or even bigger; in a city where the traffic condition is very good, we set $p$=5 or even smaller. In this paper, we set $p$=10.

we define the predict probability of congestion of road $i$ at time $t$ as $\hat{p}^{i}_{t}$, and the real probability of congestion is defined as $\hat{p}^{i}_{t}$, our problem can be formulated as:
\begin{equation}
\begin{split}
Min_{f}:Criterion(p^{i}_{t+d},\hat{p}^{i}_{t+d})\\
\textit{s.t.}\quad \hat{p}^{i}_{t+d}=f(V_{t:t+h}^{i})
\end{split}
\end{equation}

The $Criterion(\quad,\quad)$ in this problem can be defined as cross entropy, which gives measurements of the distance between our predict probability and the real likelihood. Note that the real likelihood is unknown even after time $t+d$, in practice we convert it to a binary classification problem, estimating the probability from history data with a machine learning method.

In Problem c, we model the probability distribution of speed at time $t$ on road $i$ as Gaussian distribution. Gaussian distribution is widely observed and used in real world applications. Let $X^{i}_{t}$ denote the probability distribution of speed on road $i$ at time $t$. $\sigma^{i}_{t}$ denote variance of speed on road $i$ at time $t$. Our assumption can be formulated as:
$X^{i}_t\sim\mathcal N(x^{i}_t,\sigma^{i}_{t})$.

 We use the method in Problem a to estimate $x^{i}_t$. Due to the time varying pattern of traffic flow speed data, $\sigma^{i}_{t}$ can't be assumed to be constant. In this paper, we use maximum likelihood estimation (MLE) to approximate $\sigma^{i}_{t}$ at the given time.
\begin{equation}
   \overline{x}_{t:t+h-1}^i = \frac{1}{h}\sum_{j=t}^{t+h-1}x_j^i
\end{equation}
\begin{equation}
 \hat{\sigma}_{t+h}^i=\sqrt{\frac{1}{h}\sum_{j=t}^{t+h-1}(x_j^i-\overline{x}_{t:t+h-1}^i)^2}
\end{equation}

We will use $\hat{X}_{t}^t\sim\mathcal N(\hat{x}_{t}^i,\hat{\sigma}_{t}^{i})$ as the estimation of $X^{i}_t$.
\subsection{Navigation Model}
In our navigation model, We are given the traffic topology of a city. Our goal is to find an optimal path at a given time(usually in the future) from the start point to the destination. The path we find should be fast, short, unlikely to be in congestion, very stable and not vulnerable.

The traffic topology at time $t$ can be modeled as a weighted strongly connected digraph $G_t=(V_t,E_t,S_t,L_t)$. $V_t$ is defined as the set of vertices at time $t$, $E_t$ is defined as the set of edges at time $t$, and $S_t,L_t$ is defined as a map from $E$ to $\mathbb{R}$. Place $k$ at time $t$ is modeled as a vertex $v_t^k$ in $V_t$. Road $i$ at time $t$ is modeled as an edge $e_t^i$ in $E_t$. $S_t(e_t^i)$ is defined as the speed of the road $i$ at time $t$. $L_t(e_t^i)$ is defined as the length of the road $i$ at time $t$. In case of simplicity we assume the speed of the road $i$ at time $t$ on both directions remain the same.

Even though the traffic is time-varying and changes really fast. We assume some parts of the traffic topology remain the same regardless of time in this paper. We assume $V_t = V$, $E_t = E$, $L_t = L$ for any $t$. Even the old roads may be extended or the new roads may be constructed or the old roads may be deprecated, the traffic topology will not change significantly in a quite long time.

In our navigation model,we mainly focus on the following problems. Given the predict \{speed,congestion probability,speed probability distribution\}and the start point and destination of users:
\begin{itemize}
    \item [a.] Find the top $k$ fastest path from the start point to destination.
    \item [b.] Combining the prediction results to improve the performance of our navigation model(EOPF).
    
\end{itemize}

In Problem a, we model the problem as a derivative of the shortest path finding problem in graph theory. We will use a modified Dijkstra algorithm to solve it in the following section. Note that we do not know the real road speed at the time we drive from the start point to the destination when our navigation algorithm shows us a estimated shortest path. We will use the predicted speed $\hat{S}_t$ as the input of our navigation algorithm and we will show that the output path is very close to the real optimal path using $S_t$ as input.

In Problem b, we obtain the prediction of future traffic speed, the prediction of congestion probability, the prediction of probability distribution. We estimate the probability of congestion from the start point to the end in the following way: 
\begin{equation}
    Pr(Congestion(x))=\prod_{i\in\{i_1,i_2,...,i_k\}}Pr(Congestion(e_i))\label{co:1}
\end{equation}

We use these prediction results to improve the performance of our navigation algorithm, we name the process as EOPF(Enhanced Optimal Path Finding). we  will show our algorithm in detail in the following section.

Basically, in the traditional navigation algorithm, at time $t+h$, we want to find an optimal path at time $t+d(h<d)$. However, we do not know the real traffic speed $S_{t+d}$ at $t+d$ when we are actually at time $t+h$. So we estimate the traffic speed $\hat{S}_{t+d}$ in our prediction model. Of course, the optimal path $\hat{P}_{t+d}$ found using $\hat{S}_{t+d}$ instead of $S_{t+d}$ will be worse than the actual optimal path $P_{t+d}$. To solve this problem, we use the prediction results in prediction model to improve the performance of our navigation model. Our prediction results include: $\hat{p}_{t+d}$(predicted probability of congestion at time $t+d$), $\hat{\sigma}_{t+d}$(the variance of traffic speed at time $t+d$). We also use the number of roads on the path $n$ as additional information to improve the performance of the navigation model because the more roads a path has, the more vulnerable it is.

Model without EOPF:
\begin{equation}
    \hat{P}_{t+d} = Navigate(\hat{S}_{t+d})
\end{equation}

Model with EOPF:
\begin{equation}
    \hat{P}_{t+d} = Navigate(\hat{S}_{t+d},\hat{p}_{t+d},\hat{\sigma}_{t+d},n)
\end{equation}

\section{System design and analysis}
\subsection{Pipeline}
\begin{figure*}[htbp]
\centerline{\includegraphics[scale=0.80]{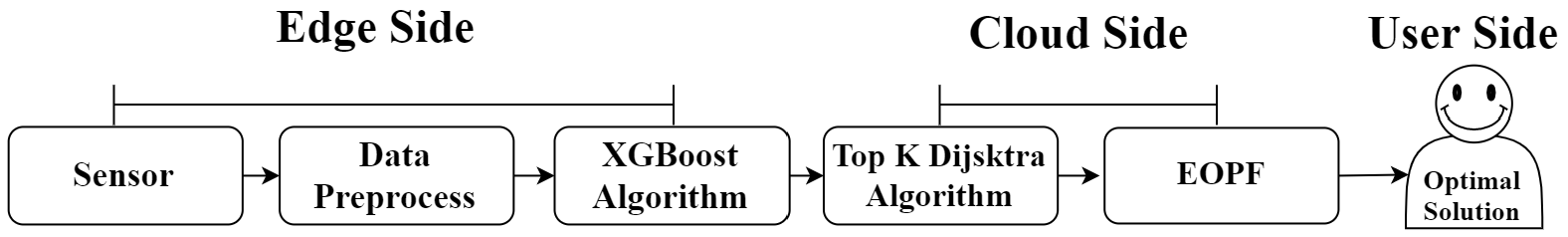}}
\caption{Pipeline of our prediction and navigation system }
\label{fig2}
\end{figure*}

Our system pipeline is shown in Fig.\ref{fig2}. Basically our model is a two-stage machine learning model. Edge devices are distributed on every road in the city. Sensors on them will collect traffic flow data periodically. The data are then sent to the data preprocess module, the noise and discontinuity in data will be cleared out and the data will be formatted and fed into our XGBoost algorithm(1st stage). Our algorithm will make predictions on future traffic conditions. The edge devices will send the prediction results and real time traffic conditions to the cloud server, which will be stored in cloud databases. When a user sends request to the cloud side for a navigation, the cloud server will run a top K Dijkstra algorithm to get a set of candidate paths. As we use $\hat{S}_{t+d}$ instead of the unknown $S_{t+d}$, the predicted shortest path may not be the real shortest path. Our 
neural network based algorithm(EOPF,2nd stage) will find one optimal path from the K paths and return the result to the user. Our neural network learns from history results of our navigation algorithm therefore it will improve the navigation output.

\subsection{XGBoost Algorithm(1st stage)}
Decision tree and tree boosting is a highly effective and widely used machine learning method. XGBoost\cite{b21} is a highly scalable and computational inexpensive tree boosting algorithm. It can be implemented on edge devices easily and is used widely by data scientists to achieve state-of-the-art results on many machine learning challenges. We will use this machine learning approach to make predictions on future traffic conditions. 

Basically, we use the state vector defined in the above section. We take the state vector $V_{t:t+h}^i$ as input feature and the real speed $x_{t+d}$ as target. We construct the dataset as $\{(V_{t:t+h}^i,x^i_{t+d})|t=0,1,2,3...n\}$, and we will split the dataset for training and testing in the following section.

A tree 
ensemble model 
uses $m$ additive functions
($F_1$,$F_2$,...$F_m$) to map the input to output. $F$ is defined as the space of decision trees(widely known as CART). The final speed prediction for a given state vector is the sum of predictions from $m$ decision trees.

\begin{equation}
    \hat{x}_{t+d}^i = f(V_{t:t+h})=\sum_{k=0}^{m}F_k(V_{t:t+h})
\end{equation}

The goal of XGBoost can be represented as an optimization problem in the following way. $T$ is the number of leaves in the tree. $\omega_k$ is defined as the weight of the leaf node at the $k$th tree. $\Omega(F_k)$ is called regularization term and it penalizes the complexity of the decision tree model. $\gamma$ and $\lambda$ are constants for regularization.
\begin{equation}
\begin{split}
     Min_{F}G = \sum_{t=0}^{n}Criterion(x_{t+d}^i,\hat{x}^i_{t+d})+\sum_{k=1}^{m}\Omega(F_k)\\
      where\quad \Omega(F_k)=\gamma T +\frac{1}{2}\lambda\|\omega_k\|^2
\end{split}
\end{equation}

The optimization problem can be explicitly solved and a highly efficient algorithm is proposed. For each road, we train one model to make predictions. So XGBoost algorithm is much faster than deep learning approach. Meanwhile, the algorithm has been implemented in many programming languages and computing platforms so that implementation on edge devices will be easy.   

XGBoost algorithm is also used to predict congestion. 
We define $y_{t+d}^i\in\{0,1\}$ is at time $t+d$ the traffic condition on road $i$ (1 for congestion,0 for no congestion) and $\hat{y}_{t+d}^i\in\{0,1\}$ is our prediction of traffic condition at time $t+d$ on road $i$. Similarly, it can be modeled as an optimization problem.
\begin{equation}
\begin{split}
     Min_{F}G = \sum_{t=0}^{n}Criterion(y_{t+d}^i,\hat{y}^i_{t+d})+\sum_{k=1}^{m}\Omega(F_k)\\
      where\quad \Omega(F_k)=\gamma T +\frac{1}{2}\lambda\|\omega_k\|^2
\end{split}
\end{equation}

\subsection{Top K Dijkstra Algorithm}
We use Yen's algorithm \cite{b20} to solve the top K shortest path problem. First, We use Dijkstra algorithm to find the shortest path from the start point $O$ to the destination $D$, defined as $P_1$. Then, Given the $i$th shortest path, to find the $i+1$th shortest path $P_{i+1}$, We define the vertices on the path other than the destination $D$ as deviation points $\{N_1,N_2,...,N_d\}$. For each deviation point $N_j$, we find its shortest path to the destination $D$. The path from the starting point $O$ to the deviation point $N_j$ is known(from $P_i$). These two paths can be concatenated to get a new path from $O$ to $D$. The new path will be added to the candidates list. The Path with the smallest cost in the candidates list will be selected as $P_{i+1}$.  In the worst case, the time complexity is $O(Kn(m+nlog(n)))$. The following process will be repeated until we find K shortest paths $P_1$, $P_2$,..., $P_k$.  
\floatname{algorithm}{Algorithm}
\renewcommand{\algorithmicrequire}{\textbf{Input:}}
\renewcommand{\algorithmicensure}{\textbf{Output:}}
\begin{algorithm}
	\caption{Top K DIJKSTRA$(G_t,O,D)$}
	\begin{algorithmic}[1]
	\REQUIRE $G_t$,$O$,$D$,$K$
	\ENSURE K shortest paths from $O$ to $D$
	\STATE{$P_1$ $\leftarrow$ getSingleShortestPathdijkstra($O$, $D$)}\\
	\STATE{$Result$.add($P_1$)}
	\STATE{$Candidates=EmptyList$}
	\STATE{$i=1$}
	\WHILE{$i<K$}
		\FOR {each vertex v on $P_i$}
			\IF{v can reach $D$}
			\STATE{$P_{v:D}$ $\leftarrow$ Subpath($v$,$D$) on $P_i$}\\
			\STATE{$P_{O:v}$ $\leftarrow$ getSingleShortestPathdijkstra($O$,$v$)}\\
			\STATE{$Candidates$.add($(P_{O:v}-v-P_{v:D})$)}
			\ENDIF
		\ENDFOR
	\STATE{$P_{i+1}$ $\leftarrow$ $Extract_Min(Candidates)$ }\\
	\STATE{$i=i+1$}\\
	\ENDWHILE
	\RETURN $Result$
	\end{algorithmic}
\end{algorithm}
\subsection{EOPF(Enhanced Optimal Path Finding,2nd stage)}
 Our goal is to find one optimal path from the K paths from the top K Dijkstra algorithm and return the result to the user. We develop an algorithm based on neural network to realize our goal.
 
Basically, our approach is still to better predict the future speed in order to get a optimal path in future traffic conditions. We collect history requests, predictions and real speed to create a new dataset $\{((\hat{S}_{t+d},\hat{p}_{t+d},\hat{\sigma}_{t+d},n),S_{t+d})|t=0,...,T \}$. We use a neural network with two hidden layers(denoted as $NN(...)$) to approximate the target output. The output of the neural network is denoted as $\hat{S}_{t+d}^{EOPF}$. We use sigmoid function as activation function and we adopt MSE(mean squared error) as our optimization objective.
\begin{equation}
    \begin{split}
    Min:\|\hat{S}_{t+d}^{EOPF}-S_{t+d}\|_2\\
        \hat{S}_{t+d}^{EOPF} = NN(\hat{S}_{t+d},\hat{p}_{t+d},\hat{\sigma}_{t+d},n)
    \end{split}
\end{equation}

\section{Evaluation}
\subsection{Dataset Description}
In this paper, We do not really use our devices or sensors to collect real time traffic data, instead, we utilize one real-world network scale traffic speed data. Our dataset is frequently used in related research and is considered to be highly representative. The dataset was collected in the highway of Los Angeles in real time by loop detectors. We get this dataset from \cite{b1}. The dataset contains data from 207 sensors from Mar.1 to Mar.7, 2012. The time interval of our dataset is 5 minutes. Meanwhile, our dataset includes an adjacency matrix. The adjacency matrix is calculated by the distance between sensors in the trafﬁc network. We use linear interpolation to fill missing values and clear out the discontinuities of the data. In our prediction algorithm, we use 80\% of the dataset for training and 20\% of the dataset for testing.
\subsection{Environment Setup}
Our data preprocess algorithms are implemented with Python 3.6.8 and Numpy 1.17.4 package. The XGBoost algorithm is implemented by XGBoost 0.90 package in Python. The EOPF algorithm is implemented by scikit-learn 0.22.1 package in Python. The top K Dijkstra algorithm is implemented in Java to achieve better performance. We run all our experiments with PC which has a Intel core i7-8650U CPU@1.90GHz and 16GB RAM. 
\subsection{Evaluation Of Prediction Algorithms}
\subsubsection{Evaluation Metrics}
In this paper, we use 3 metrics to evaluate the performance of our XGBoost speed prediction algorithm,including root mean squared error(RMSE) 

\begin{equation}
    RMSE_t=\sqrt{\frac{1}{n}\sum_{j=0}^{n}(\hat{x}_{t+j}^{i}-x_{t+j}^{i})^2}
\end{equation}

mean absolute error(MAE)
\begin{equation}
   MAE_t=\frac{1}{n}\sum_{j=0}^{n}\|\hat{x}_{t+j}^{i}-x_{t+j}^{i}\|
\end{equation}

accuracy(Acc)
\begin{equation}
    Acc = 1-\frac{\|\hat{x}^i-x^i\|}{\|x^i\|}
\end{equation}

Meanwhile, 
we use precision
(positive predictive value),
recall(sensitivity)
and F1-score(average of precision and recall) to evaluate the performance of our XGBoost congestion prediction algorithm.

\subsubsection{Baseline Models}
We use 8 baseline models for the speed prediction problem and 5 baseline models for the congestion prediction problem.Our baseline models include:
\begin{itemize}
    \item History Average: Take the average of history speed data.
    \item SVM: Support Vector Machine.
    \item ARIMA: Auto-Regressive Integrated Moving Average model.
    \item Neural Network: with two hidden layers.
    \item KNN:K-Nearest Neighborhood model
    \item Model blending and Weighted Average: SVM ensembles with Neural Network.
\end{itemize}
\subsubsection{Comparison With Baseline Models}
\begin{table}[]
\caption{Comparison with baseline models(Speed prediction)}
\center
\begin{tabular}{|l|l|l|l|l|}

\hline
Model/Metrics & RMSE            & MAE             & Acc            & Time(s)            \\ \hline
History Average              & 6.8587          & 3.6875          & 0.8833          & 0.0000          \\ \hline
SVR                          & 4.3668          & 2.5775          & 0.9257          & 1.4785          \\ \hline
Neural Network               & 4.4029          & 2.6831          & 0.9251          & 1.8264          \\ \hline
KNN-Uniform                  & 4.9370          & 2.8879          & 0.9160          & 0.0249          \\ \hline
KNN-Weighted                 & 4.9108          & 2.8774          & 0.9164          & 0.0237          \\ \hline
ARIMA                        & 10.0423         & 7.5557          & 0.8277          & 0.1737          \\ \hline
Model Blending               & 4.5849          & 2.8020          & 0.9220          & 2.3540          \\ \hline
Weighted Average             & 4.3482          & 2.6105          & 0.9260          & 1.9672          \\ \hline
XGBoost                      & \textbf{4.2915} & \textbf{2.5784} & \textbf{0.9270} & \textbf{1.0359} \\ \hline
\end{tabular}
\label{tab1}
\end{table}
\begin{table}[]
\center
\caption{Comparison with baseline models(Congestion prediction)}
\begin{tabular}{|l|l|l|l|l|}
\hline
Model/Metrics    & Precision    & Recall        & F1-score      &  Time(s)        \\ \hline
SVC              & 0.86         & 0.57          & 0.68          & 1.7886          \\ \hline
Neural Network   & 0.83         & 0.46          & 0.59          & 1.8890          \\ \hline
KNN-Uniform      & 0.85         & 0.58          & 0.69          & 0.0252          \\ \hline
KNN-Weighted     & 0.82         & 0.61          & 0.70          & 0.0279          \\ \hline
\textbf{XGBoost} & \textbf{0.80} & \textbf{0.65} & \textbf{0.72} & \textbf{1.1359} \\ \hline
\end{tabular}
\label{tab2}
\end{table}
We train the XGBoost algorithm for 50 rounds and compare the performance with baseline models. In TABLE \ref{tab1}, we show that XGBoost outperforms all other models in RMSE and Acc. Meanwhile, the XGBoost trains and predicts relatively fast in comparison with other machine learning methods. We also find that ARIMA performs poorly in this experiment. One reason for that is ARIMA is designed for stationary series, but traffic speed data is highly non-stationary and time-varying. It can not work well on large scale real time traffic speed data. SVM and neural network seem to be really effective methods on that task, but they are a little slower than XGBoost. Ensemble learning methods like model blending and weighted average perform well, but they are not easy to implement in comparison with the highly scalable XGBoost model. KNN method gives us a surprise: though the algorithm is quite simple, its performance is not bad considering the time cost. In TABLE \ref{tab2}, we show that XGBoost outperforms all other models in recall and F1-score. A higher recall means that more congestion situations are predicted and warned so in practice users can avoid more possible congestion. Though the precision of some other methods are high, they fail to predict many possible congestion situations. Therefore they perform worse than XGBoost.

\subsubsection{Further Evaluation}
In this part, We design several experiments to better show the performance of our model.
\begin{itemize}
    \item how time interval is related to performance.
    
    Intuitively, if we want to make predictions on the traffic condition 5 minutes later, we will predict very accurately since the time we predict is very close to the time we are now at, and the traffic condition won't change much. However, if we make predictions on the traffic conditions 1 hour later, we can hardly predict the traffic condition because the traffic condition varies very fast and unexpected accidents may happen during that time. From Fig.\ref{fig4} and TABLE \ref{tab3}, we show that with the increase of time interval, our prediction accuracy drops and error becomes larger. With a time interval of 30 minutes, our prediction algorithm preforms poorly, and we should remember to avoid predicting the traffic condition for a long time.
    \begin{figure}[htbp]
    \centerline{\includegraphics[scale=0.90]{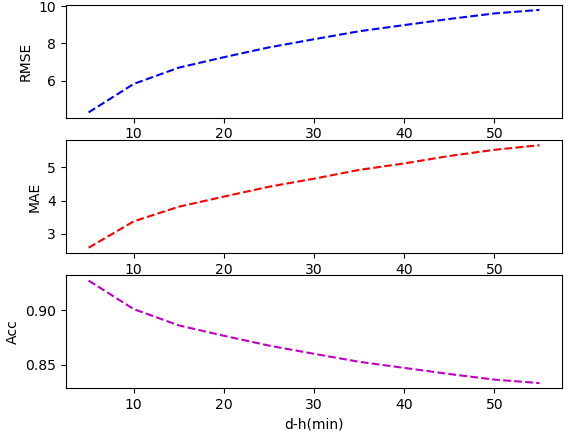}}
    \caption{The relationship between time interval and performance}
\label{fig4}
\end{figure}
\begin{table}[]

\center

\caption{Time interval and performance comparison}
\label{tab3}
\begin{tabular}{|c|c|c|c|}
\hline
Interval/Metrics  & RMSE            & MAE             & Acc             \\ \hline
5min           & 4.2915          & 2.5784          & 0.9270          \\ \hline
10min          & 5.8214          & 3.3734          & 0.9009          \\ \hline
15min          & 6.6960          & 3.8129          & 0.8860          \\ \hline
20min          & 7.2571          & 4.1180          & 0.8765          \\ \hline
25min & 7.7852 & 4.4155 & 0.8675 \\ \hline
30min          & 8.2226          & 4.6574          & 0.8600          \\ \hline
35min          & 8.6519          & 4.9226          & 0.8527          \\ \hline
40min          & 8.9823          & 5.1163          & 0.8471          \\ \hline
45min          & 9.3121          & 5.3409          & 0.8415          \\ \hline
50min          & 9.6083          & 5.5262          & 0.8364          \\ \hline
55min          & 9.7996          & 5.6644          & 0.8332          \\ \hline
\end{tabular}
\end{table}
    \item how performance is related to training epoches.\\
    In machine learning, training more epoches means less error and more accurate results, however, too much training is often associated with overfitting. In Fig.\ref{fig3}, we show that after training nearly 50 epoches, performance on the test set stops improving. After training 100 epoches, performance on the training set stops improving. So we can stop after training 50 epoches. Another interesting thing is that test error is always significantly larger than training error, which is the sign of overfitting. One reason for that is the traffic speed data is highly non-stationary and time varying so our algorithm may learn from useless patterns in data.
    \begin{figure}[htbp]
    \centerline{\includegraphics[scale=0.25]{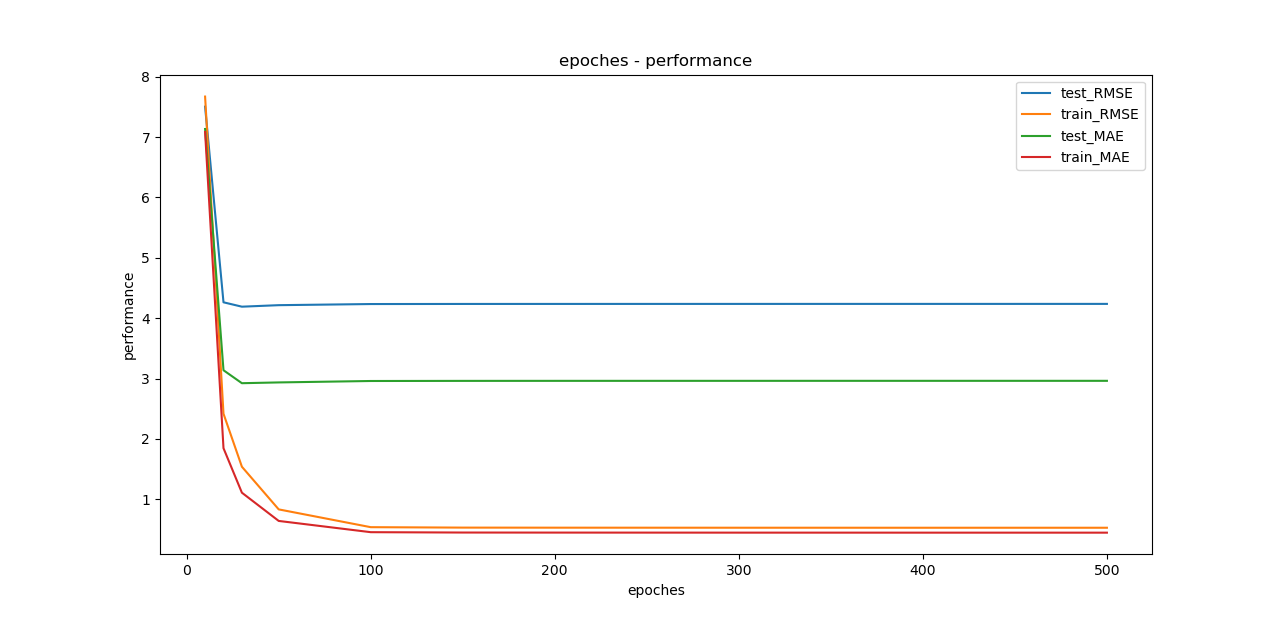}}
    \caption{The relationship between training epoches and performance}
\label{fig3}
\end{figure}
    
    \item the ROC curve of congestion prediction.\\
     ROC stands for Receiver Operating Characteristic\cite{b19}. Currently, ROC curves are often used to see how our predict model can distinguish between the true positives and negatives. AUC is defined as the area between ROC curve and x-axis, and its value is between 0 and 1. A larger AUC means a better predict model. The AUC value of our congestion prediction model is 0.932, which means that our congestion prediction algorithm works quite well. The ROC curve of our model is shown in Fig.\ref{fig5}.

    \begin{figure}[htbp]
\centerline{\includegraphics[scale=0.40]{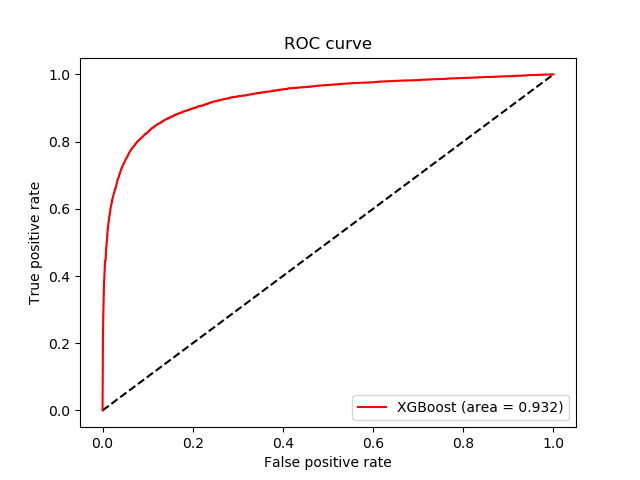}}
\caption{ROC curve of our congestion prediction algorithm}
\label{fig5}
\end{figure}

\end{itemize}
\subsubsection{Usability Study}
\begin{figure}[htbp]
\centerline{\includegraphics[scale=0.40]{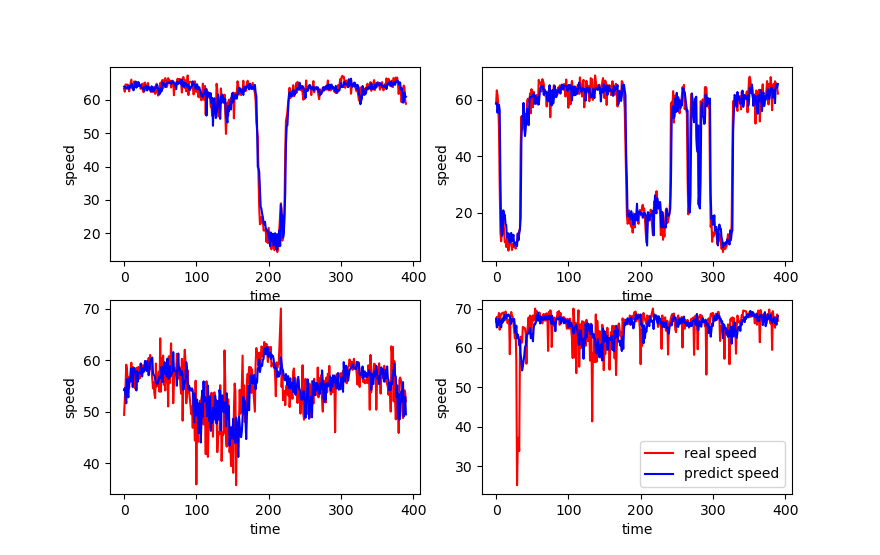}}
\caption{Different performances at several situations}
\label{fig6}
\end{figure}

Our model doesn't always perform well on the dataset, which is also the problem of many other prediction algorithms. We randomly select 4 roads(their road IDs are 69,37,18,44) and plot our predictions in blue line in Fig.\ref{fig6}. We also plot real traffic speed data in red line in Fig.\ref{fig6}.

In Fig.\ref{fig6}, we show that our algorithm performs well at road 69 and road 37. Our algorithm captures the trend of speed at road 18, but the prediction is not accurate and many details are lost. Our algorithm fails to capture the trend of speed at road 44, the speed changes quite drastically but our prediction is relatively smooth.

These results show that our algorithm performs well when traffic condition doesn't change drastically(at road 69,37). The algorithm performs poorly at peak values(at road 18,44). The algorithm can capture the trend of speed change when traffic condition doesn't change in high frequency(at road 69,37,18).

We propose 2 factors that affect our model performance and usability. The first factor measures the frequency of speed change. We use Fast Fourier Transform(FFT)\cite{b17}\cite{b18} and analysis the speed data in frequency domain. 
\begin{equation}
    FFT(x^i)[k]=\sum_{t=<N>}x_t^ie^{jt\frac{2\pi}{N}k}(k = 0,1,...N-1)
\end{equation}
\begin{table}[]
\center
\caption{Quantitatve analysis of model usability}
\label{tab4}

\begin{tabular}{|c|c|c|c|}
\hline
K(SR)  & RMSE    & MAE     & Acc     \\ \hline
1  & -0.6271 & -0.6472 & 0.7212  \\ \hline
2  & -0.629  & -0.6491 & 0.7229  \\ \hline
3  & -0.6323 & -0.6524 & 0.7257  \\ \hline
5  & -0.6385 & -0.6603 & 0.7305  \\ \hline
10 & -0.6711 & -0.692  & 0.7464  \\ \hline
15 & -0.6889 & -0.7137 & 0.7628  \\ \hline
20 & -0.6987 & -0.7277 & 0.7715  \\ \hline
K(Bias)  & RMSE    & MAE     & Acc     \\ \hline
1  & 0.4379  & 0.4539  & -0.604  \\ \hline
2  & 0.4539  & 0.4685  & -0.605  \\ \hline
3  & 0.4506  & 0.4658  & -0.5986 \\ \hline
5  & 0.4662  & 0.4777  & -0.6029 \\ \hline
10 & 0.4821  & 0.4808  & -0.61   \\ \hline
15 & 0.4809  & 0.4831  & -0.6103 \\ \hline
20 & 0.4882  & 0.4899  & -0.6156 \\ \hline
\end{tabular}

\end{table}
\begin{figure}[htbp]
\centerline{\includegraphics[scale=0.30]{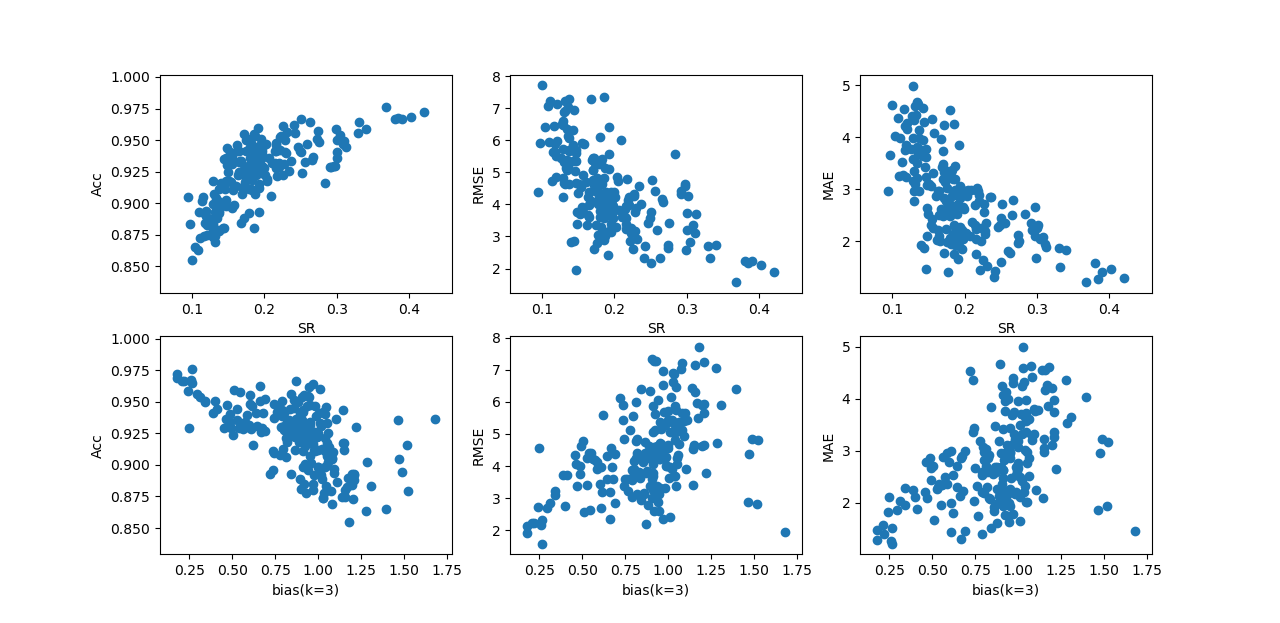}}
\caption{Quantify the model usability}
\label{fig7}
\end{figure}
Intuitively, if the spectrum of speed data has more high frequency part, the speed will change more drastically and frequently, and it will be harder to make accurate predictions; if the spectrum of speed data has more low frequency part, the speed will be smoother, and our predictions will be more accurate. We define $SR^i = \frac{\|FFT(x^i)[0:K]\|_1}{\|FFT(x^i)[0:N-1]\|_1}$, and $FFT(x^i)[0:K]=(FFT(x^i)[0],FFT(x^i)[1],...,FFT(x^i)[K])$. $SR$ measures the low frequency proportion in speed data. In the first row of Fig.\ref{fig7}, we show that $SR$($K$=5) is highly positively related to prediction accuracy and highly negatively related to prediction error. We also show in TABLE \ref{tab4} that for different $K$, the relationship remains the same.

The second factor measures the peak values of speed data. Intuitively, if the range of speed is very large, it will be hard to predict accurately of future speed; if the range of speed is relatively small, it is easier for predictions. We define $Bias^i = \frac{Max_K(x^i)-Min_K(x^i)}{mean(x^i)}$. $Max_K(x^i)$ is the $K$th largest value of speed on road $i$, $Min_K(x^i)$ is the $K$th smallest value of speed on road $i$ and $mean(x^i)$ is the average speed on road $i$. In the second row of Fig.\ref{fig7}, we show that $Bias$($K$=3) is highly negatively related to prediction accuracy and highly positively related to prediction error. We also show in TABLE \ref{tab4} that for different $K$, the relationship remains the same.

In practice, we can use these 2 factors as indicators of poor performance of our algorithm. Before using our model on real roads, we can computer $SR$ and $Bias$ on these roads. If $SR$ of the road is low and $Bias$ on the road is high, we can say that the traffic condition of the road is full of unexpectation and highly time varying, and we can not make accurate predictions on the traffic condition. We can warn the users to avoid driving on these roads and we can try to design special algorithms for that kind of roads in the future.
\subsection{Evaluation Of Navigation Algorithms}
We use the same traffic dataset as the prediction algorithms. Our dataset includes data from 391 time intervals and we generate 100 random user requests from any start point to the end $(O,D)$ for each time interval. Our results show that using the predict speed $\hat{S}_t$ as input, the output path can be at most 1.6\% slower that the real optimal path on average, which is acceptable in most cases. In more than 66\% of the time, the output path is exactly the real optimal path. Moreover, after using the EOPF algorithm,
the performance of our navigation algorithm can improve by over 7\% on average on the entire dataset, and in the situations where the performance is improved, the average improvement is more than 8\%.
\subsubsection{Without EOPF}We compare the performance of our model with the model where we don't use any prediction algorithm and just use the current speed as the "predict speed" of future traffic conditions. In this part we don't use the EOPF method. In TABLE.\ref{t6}, we show that in general using the predict speed to find the optimal path is much better than using the current speed as alternative, which shows the effectiveness of our prediction algorithm. When the time interval is short, our algorithm performs much better than the naive approach. However, when the time interval becomes longer, both methods perform poorly. One possible reason for that is the traffic speed is highly time varying so that attempts trying to predict further traffic conditions are very hard.
\begin{table}[]
\caption{Comparison with naive method}
\label{t6}
\center
\begin{tabular}{|c|c|c|}
\hline
Time  & Naive  & Predict \\ \hline
5min  & 0.92\% & 0.49\%  \\ \hline
10min & 2.96\% & 2.58\%  \\ \hline
15min & 3.26\% & 4.13\%  \\ \hline
20min & 5.92\% & 5.91\%  \\ \hline
25min & 8.71\% & 7.99\%  \\ \hline
\end{tabular}
\end{table}
\subsubsection{EOPF Performance}
After using EOPF as an approach to enhance the performance of our model, in many situations the performance will be improved. Though in some cases, using EOPF will entail a worse result, the overall average improvement is still promising. We find that after using EOPF method, the output of 286 out of 9100 requests are improved. The average improvement is 8\%, which will be a greater number on a larger dataset. However, there are 273 out of 9100 requests whose outputs become worse. The average improvement on the entire dataset. In TABLE.\ref{t8}, we show the distribution of the improvement, which shows that even there are some cases our method doesn't work well, in more cases our performance will be improved.

\begin{table}[]
\caption{Performance after using EOPF(compared with non-eopf method)}
\label{t8}
\center
\begin{tabular}{|c|c|c|}
\hline
Method               & Better  & Worse   \\ \hline
\textgreater{}30\% & 1.40\%  & 0.70\%    \\ \hline
20\%-30\%          & 6.29\%  & 0.51\%    \\ \hline
10\%-20\%          & 23.78\% & 16.12\%  \\ \hline
5\%-10\%           & 22.73\% & 23.11\%   \\ \hline
\textless{}5\%     & 45.80\% & 55.56\%  \\ \hline
\end{tabular}
\end{table}

We also compare the output of our model with the real optimal path. Of course, our output is always worse than the real optimal path. In TABLE.\ref{t9}, we show that After using EOPF method, the performance of model improves by 7\% in comparison with the model without EOPF. Before using EOPF, our model is 1.62\% worse than the real optimal path on average, and our model is 1.51\% worse than the real optimal path on average. The cases when our model is 20\% worse than the real optimal path are significantly avoided. Though this improvement is small, it can be really useful when used in practice on large scale data.
\begin{table}[]
\caption{Performance after using EOPF(compared with optimal path)}
\label{t9}
\center
\begin{tabular}{|c|c|c|}
\hline
Method             & EOPF    & Without EOPF \\ \hline
\textgreater{}30\% & 0.38\%  & 0.56\%       \\ \hline
20\%-30\%          & 0.89\%  & 1.12\%       \\ \hline
10\%-20\%          & 3.23\%  & 3.22\%       \\ \hline
5\%-10\%           & 5.21\%  & 4.80\%       \\ \hline
\textless{}5\%     & 90.29\% & 90.30\%      \\ \hline
Overall            & 1.51\%  & 1.62\%       \\ \hline
\end{tabular}
\end{table}

\section{Conclusions}
In this paper, we propose a two-stage machine learning approach to enhance the performance of the navigation system. In the first stage, we show that XGBoost algorithm can perform quite well in traffic speed prediction and congestion prediction, and it can outperform many other methods both in the accuracy of prediction and the cost of time. Additionally, we propose two factors to quantify the usability of our model and these two factors can be considered as measurements for the difficulty of time series prediction. Between the two stages, we use the top K Dijkstra algorithm as the basis of our navigation system, which gives K shortest paths on the predict data as candidates for the optimal path. In the second stage, we show that our neural network based EOPF method can efficiently find one optimal path form K shortest paths in comparison with the naive method and the method without EOPF. Our method shows significant improvement (7\%) on the dataset on average, which will be really powerful in real applications.

There are many topics future researchers can do in this field. Predicting the future traffic conditions will continue to be a hot topic and more accurate prediction methods will be developed. Using machine learning models to enhance the performance of navigation system is a great topic in real applications. Future research including factors on traffic management will be quite promising.

\vspace{12pt}

\end{document}